\documentclass[fleqn,usenatbib]{mnras}

\usepackage[T1]{fontenc}

\DeclareRobustCommand{\VAN}[3]{#2}
\let\VANthebibliography\thebibliography
\def\thebibliography{\DeclareRobustCommand{\VAN}[3]{##3}\VANthebibliography}

\usepackage{graphicx}	
\usepackage{amsmath}	
\usepackage{amssymb}	

\title[Stellar wind and efficiency of plasma radio source]{Effect of stellar wind on the efficiency of plasma radio emission from exoplanet {HD189733b}}

\author[V. V. Zaitsev et al.]{
V.V. Zaitsev,$^{1,2}$
V.E. Shaposhnikov,$^{1,3}$\thanks{E-mail: sh130@ipfran.ru}
M.L. Khodachenko,$^{4}$
and M.S. Rumenskikh$^{5,6}$
\\
$^{1}$Institute of Applied Physics of the Russian Academy of Sciences, Nizhny Novgorod,
Russia\\
$^{2}$Pulkovo Observatory of Russian Academy of Sciences, Saint-Petersburg, Russia\\
$^{3}$High School of Economics, National Research University, Nizhny Novgorod Branch, Nizhny Novgorod, Russia\\
$^{4}$Space Research Institute? Austrian Academy of Sciences, Graz, Austria\\
$^{5}$Institute of Laser Physics SB RAS, Novosibirsk, Russia\\
$^{6}$Institute of Terrestrial Magnetism, Ionosphere and Radio Wave Propagation of the
Russian Academy of Sciences,  Troitsk, Russia
}

\date{Accepted XXX. Received YYY; in original form ZZZ}

\pubyear{2023}

\begin{document}
\label{firstpage}
\pagerange{\pageref{firstpage}--\pageref{lastpage}}
\maketitle

\begin{abstract}

On the example of the exoplanet HD 189733 an influence of stellar activity on the efficiency of the plasma mechanism of radio emission generation of the exoplanet and the properties of this emission are considered. The plasma generation mechanism can be effectively implemented in the plasmasphere of exoplanets with a weak magnetic field and a relatively high electron plasma density, when the electron cyclotron maser is not efficient. The plasma  mechanism depends essentially on the parameters of the plasma and involves the generation of plasma waves by energetic electrons and a conversion of these waves into electromagnetic radiation. The stellar wind can significantly modify the exoplanet plasmasphere, which was not taken into account in the first studies of the plasma mechanism in the plasmasphere of HD 189733b. In present  study we used a three-dimensional model of the interaction of the exoplanet HD189733b with the stellar wind for cases of moderate and intense stellar winds.  The study shows that the implementation of plasma mechanism is possible at any intensity of the stellar wind. However, depending on the intensity, the requirements for the parameters of plasma mechanism change. In particular, the plasma waves energy which is required to generate the radio emission available for registration by modern radio telescopes changes. Besides, the frequency range of the radio emission changes. The latter will make it possible to use the detected radio emission as an indicator of the activity of the parent star.

\end{abstract}




\section{Introduction}

Currently, the study of exoplanets and exoplanetary systems is one of the important areas in astrophysics. The study of exoplanets in the visible frequency range is hampered by the high contrast between the luminosity of the star and the planet, as well as their relatively small angular separation. In the radio range, the situation is different, here the contrast is much lower and it is possible to separate the exoplanet's own radiation from the radiation of its central star \citep{Zarka(2007)}.
Therefore, the question arises of what parameters the atmosphere of an exoplanet should have in order for its radio emission to be detected by modern radio astronomical equipment   \citep{Griessmeier(gzs)(2007),Zarka(2007),Jardine(jc)(2008),Nichols(nm)(2016),Burkhart(bl)(2017),Vidotto(vd)(2017),Weber(wlsckgrvmok)(2017),Weber(weiogflr)(2018),
Selhorst(sbcsvv)(2020),Turnpenney(tnwb)(2018),Turnpenney(tnwj)(2020),Narang(nmilhtmum)(2021),Narang(nohmbt)(2023)}. Usually, when answering this question, the mechanism of radio emission generation based on the instability of electrons at the cyclotron frequency (electron cyclotron maser \citep{Wu(wl)(1979)}) in the plasmasphere of  exoplanet is considered. This mechanism is effective in  sufficiently strong magnetic field, when the cyclotron frequency significantly exceeds the plasma frequency  \citep{Melrose(mdh)(1984)}. Moreover, \cite{Weber(wlsckgrvmok)(2017),Weber(weiogflr)(2018)} show that this mechanism is only effective in supermassive exoplanets with strong magnetic fields.  On exoplanets with a mass on the order of the mass of Jupiter, this mechanism is not effective due to the presence of an extended and relatively dense plasma shell or a relatively weak magnetic field. Therefore, they conclude that exoplanets with a weak magnetic field, such as, for example, HD 209458b and HD 189733b, do not have own radio emission.

On the other hand, the plasma mechanism does not require a strong magnetic field to efficiently generate radio emission \citep{Zaitsev(zs)(1983)}. ). The model based on the plasma mechanism of radio emission involves the generation of plasma waves by energetic electrons and their conversion into electromagnetic radiation. The conversion can be at the plasma frequency upon scattering by plasma particles or at the double plasma frequency as a result of Raman scattering (coupling of two plasma waves). During conversion at the plasma frequency, the maser effect can occur. This effect manifests itself in an exponential increase in the intensity of electromagnetic radiation with an increase in the energy of plasma waves.  In the plasma model, the possible frequency interval of the radio emission emerging from the source is determined by the density and distribution of the plasma in the source, and not by the magnetic field. The efficiency of plasma mechanisms has been demonstrated in the theory of solar radio emission, radio emission from Jupiter, stars of late spectral classes, and neutron stars\citep{Zaitsev(zs)(1983),Zaitsev(zs)(2017),Zheleznyakov(zzz)(2012),Shaposhnikov(slzzk)(2021)}. The study of the plasma mechanism of exoplanet radio emission will allow to  both indicate new targets for search and advance in the study of the physical conditions in the plasmaspheres of exoplanets with a relatively weak magnetic field. Therefore, it seems relevant to analyze the possibility of implementing plasma mechanisms for generating radio emission in the plasmaspheres of exoplanets. On example HD 189733b  \citet{Zaitsev(zs)(2022)} showed that on exoplanets with a weak magnetic field the  plasma mechanism of radio emission can be effectively implemented instead of the electron cyclotron maser. Using the simple plasmosphere model by  \citet{Guo(2011)} that does not take into account the effect of the stellar wind \citet{Zaitsev(zs)(2022)} determined
 the frequency range of possible radio emission and the requirements for the energy density of plasma waves, at which the intensity of radio emission is sufficient for observation by modern radio telescopes. Note that the stellar wind significantly affects the radio emission fluxes from exoplanets with a strong magnetic field, where the main mechanism of radio emission generation is the electron cyclotron maser \citep{Griessmeier(gpkmmr)(2007)}.

\citet{Rumenskikh(rsklmbf)(2022)} showed that the stellar wind can significantly modify the exoplanet's plasmasphere. Since the efficiency of the plasma generation mechanism depends significantly on the plasma parameters, the stellar wind can also affect the radio emission fluxes from exoplanets with a weak magnetic field.  In this paper, we study the influence of stellar activity, expressed in terms of the intensity of the stellar wind, on the properties of plasma radio source at the exoplanet HD 189733b. We investigate the properties of radio emission at both the plasma frequency and the double plasma frequency, the frequency ranges and the requirements for the energy density of plasma waves necessary to be able to detect radio emission by modern radio telescopes, depending on the intensity of the stellar wind. In Section 2, based on the work by \citet{Rumenskikh(rsklmbf)(2022)}, we discuss changes in the distributions of the plasma density and temperature in the plasmasphere of the exoplanet HD 189733b depending on the intensity of the stellar wind. Section~3 gives the basic equations needed for the analysis of the plasma mechanism of radio emission at the fundamental frequency and at the second harmonic of the plasma frequency. Section 4 contains calculations of the plasma mechanism parameters, radio emission frequency spectra, and plasma wave energies at various stellar wind intensities . Section~5 contains a discussion of the results and main conclusions.
  
  \section{Effect of the stellar wind on the plasmasphere parameters of {HD 189733b}}

Below are some parameters of the exoplanet {HD 189733b} - the distance to the Earth $R_{\rm so}\approx 63$~light years, the radius of the planet $R_{\rm p}\approx 1.14R_{\rm J}$  and its mass $M_{\rm p}\approx 1.13M_{\rm J}$. These parameters are close to the corresponding values of $R_{\rm J}$ and $M_{\rm J}$ for Jupiter, as well as estimates of the effective temperature, $T_{\rm eff}\approx 1200 \mbox{ K}$, and magnetic field near the planet's surface, $B\approx 1.8 \mbox{ G}$ \citep{Griessmeier(gzs)(2007)}. The parent star, around which the exoplanet orbits at a distance of three hundredths of an astronomical unit, is a yellow dwarf located in the constellation of Vulpecula. It has a size and temperature close to those of the Sun.

\begin{figure}
	\includegraphics[width=\columnwidth]{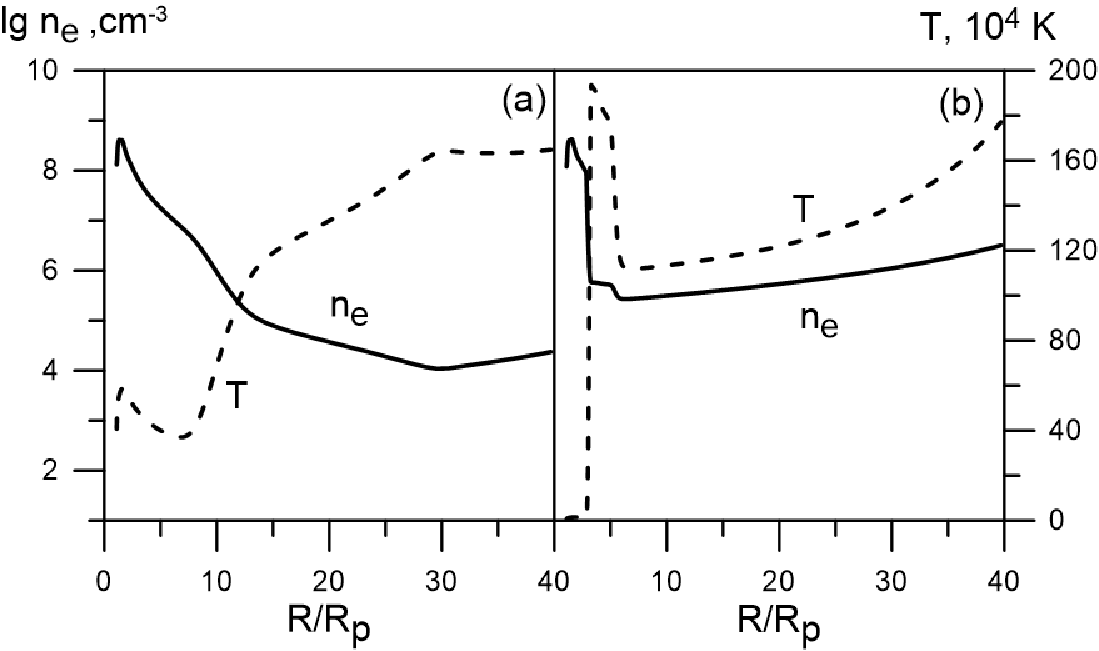}
    \caption{Distributions of plasma density $n_{\rm e}$ and temperature $T$ for a three-dimensional model of the interaction of the exoplanet HD~189733b with the stellar wind of the parent star at a) moderate wind intensity (the star mass carried away is
    $10^{11} \mbox{ g s$^{-1}$}$) and b) at an intense wind (the star mass carried away is $2 \times 10^{13} \mbox{ g s$^{-1}$}$).  }
    \label{fig1}
\end{figure}

\citet{Rumenskikh(rsklmbf)(2022)} developed a self-consistent multi-fluid hydrodynamic model of the atmosphere of hot Jupiter HD189733b on the basis of an analysis of the evolution of the spectral lines of hydrogen and helium during the transit of an exoplanet through the disk of a star.  Besides, they  simulated possible variants of stellar activity and its impact on exoplanet plasmasphere. Fig.~\ref{fig1} shows the distributions of the plasma density n and temperature $T$ for a three-dimensional model of the interaction of the exoplanet HD~189733b with the stellar wind at a moderate wind (the star mass carried away is $10^{11} \mbox{ g s$^{-1}$}$, Fig.~\ref{fig1}a) and at an intense wind (the star mass carried away is $2\times10^{13} \mbox{ g s$^{-1}$}$, Fig.~\ref{fig1}b). In this and subsequent figures, all distances are measured from the center of the exoplanet.

It follows from Fig.~\ref{fig1} that with an increase in the intensity of the stellar wind flow, a significant change in the structure of the plasmasphere occurs - a sharp decrease in the plasma density by approximately two orders of magnitude, starting from a distance of $\approx 1.5$ radii from the surface, turns into a ''plateau'' at large distances, and the temperature jump by two orders of magnitude also at a distance of about 1.5 radii. Thus, given that the efficiency of the plasma mechanism depends significantly on the plasma density and temperature, the activity of the star should affect the frequency range and intensity of the radio emission of the exoplanet. The analysis of this problem is the purpose of our work.

\section{Generation of plasma waves in the weakly magnetized plasmasphere of HD~189733b}

On exoplanets with a weak magnetic field, when the electron gyrofrequency $\omega_{\rm c}=eB/m_{\rm e} c$ is much less than the plasma frequency
$\omega_{\rm L}=\sqrt{4\pi e^2 n/m_{\rm e}}$    ($n$ is the plasma density, $B$ is the magnetic field, $e$ and $m_{\rm e}$ is the electron charge and mass,  $c$ is the speed of light), so that the condition $\omega_{\rm c}<< \omega_{\rm L}$ is satisfied, the plasma mechanism of radio emission can be effectively implemented. The mechanism involves the generation of plasma waves by energetic electrons in the source and their transformation into radio emission at the plasma frequency or at the double plasma frequency.

The plasma model assumes that radio emission is generated in a region of the planet's plasmasphere filled with an equilibrium weakly anisotropic plasma, $\omega_{\rm c}\ll \omega_{\rm L}$, with a density $n$ and temperature $T$ and with a small admixture of nonequilibrium energetic electrons with a density $n_{\rm s}\ll n$ and temperature $T_{\rm s}\gg T$. Due to the nonequilibrium electrons, plasma (Langmuir) waves are excited with a frequency
\begin{equation}
\omega^2_{\rm p}=\omega^2_{\rm L}+3k^2v^2_{\rm T},
\label{om_p}
\end{equation}
where $\vec{k}$ is the wave vector of plasma waves, $v_{\rm T}$ is the thermal velocity of electrons in the main plasma. For definiteness, we will simulate the velocity distribution of energetic electrons by a function with a loss cone
\begin{equation}
f_{\rm s}=(v_\parallel, v_\perp)=\frac{n_{\rm s}}{4\sqrt{(2\pi)^3}}\left(\frac{m_{\rm e}}{\kappa_{\rm B}T_{\rm s}}\right)^{5/2}v^2_\perp\exp\left[-\frac{m_{\rm e}\left(v^2_\parallel+v^2_\perp \right)}{2\kappa_{\rm B}T_{\rm s}}\right],
\label{f_s}
\end{equation}
where $v_\parallel, v_\perp$ are the longitudinal and transverse components of the velocity vector of energetic electrons with respect to the direction of the magnetic field in the region of plasma wave generation, $\kappa_{\rm B}$ is the Boltzmann constant. This function provides the necessary anisotropy associated with the deficit of electrons with low transverse velocities. The grow rate of the loss cone instability  has the following form \citep{Zaitsev(zs)(1975)}
\begin{multline}
\gamma(\omega, k) =  \frac{1}{4}\sqrt{\frac{\pi}{2}}\frac{\omega_{\rm p}^4}{k^3v_{\rm Ts}^3}\frac{n_{\rm s}}{n}\frac{\omega_{\rm p}^2 k^{-2}v_{\rm Ts}^{-2}+2\cot^2\alpha-1}{1+\cot^2\alpha} \times  \\
   \exp \left(-\frac{\omega_{\rm p}^2}{2k^2v_{\rm Ts}^2}\right),
\label{incr}
\end{multline}
where $v_{\rm Ts}=\sqrt{\kappa T_{\rm s}/m_{\rm e}}$ is the characteristic velocity of hot electrons. According to (\ref{incr}), the grow rate is negative ($\gamma<0)$), i.e. the system is unstable with respect to perturbations for plasma waves with phase velocities $v_{\rm ph}$
\begin{equation}
v_{\rm ph}^2<v_{\rm Ts}^2(1-2\cot^2\alpha),
\label{v_ph}
\end{equation}
which propagate at angles $\alpha >\alpha_{\rm cr}=\arctan \sqrt{2}$ with respect to the magnetic field ($\alpha$ is the  angle between   $\vec{k}$
   and  $\vec{B}$). The maximum value of the growth rate is
\begin{equation}
                           \left|\gamma_{\max} \right|\approx 3 \times 10^{-2} \frac{n_{\rm s}}{n}\omega_{\rm L}
  \label{incr_max}
  \end{equation}
for waves propagating almost orthogonally to the magnetic field, $\alpha\approx \pi/2$, with phase velocities $v_{\rm ph}\approx 0.74 v_{\rm Ts}$  and wave numbers $|\vec{k}_{\rm opt}|= k_{\rm opt} \approx 1.35\omega_{\rm L}/v_{\rm Ts}$.  The grow rate decreases by a factor of three when deviating from the optimal value to $k_1\approx 1.04\omega_{\rm L}/v_{\rm Ts}$   and  $k_1\approx 2.5\omega_{\rm L}/v_{\rm Ts}$, as well as when deviating by an angle of $\pm 30^\circ$ from $\approx \pi/2$. In the case of loss cone instability, the spectrum of plasma waves is axially symmetric with respect to the magnetic field. Therefore, the following formula is valid for the spectral volume
\begin{equation}
\left(\Delta \vec{k}\right)^3=2\pi\int\int k^2\sin \alpha dkd\alpha,
\label{sp-vol}
\end{equation}
where integration is performed over the entire interval of wave numbers $k$ of interacting plasma waves and the interval of angles between $\vec{k}$ and the magnetic field $\vec{B}$. As a result, for the spectral volume (\ref{sp-vol}) of plasma waves, we obtain the estimate
\begin{equation}
\left(\Delta \vec{k}\right)^3=2\pi k^2_{\rm opt}(k_2-k_1)\int_{\alpha_{\min} }^{\alpha_{\max} }\sin\alpha d\alpha\approx\xi\frac{\omega^3_{\rm L}}{c^3} .
\label{sp-vol-est}
\end{equation}
In formula (\ref{sp-vol-est}), the value
\begin{equation}
\xi=11.5\frac{c^3}{v^3_{\rm Ts}}
\label{ksi}
\end{equation}
depends on the characteristic velocity of energetic electrons that excite plasma waves. For example, with $v_{\rm Ts}=c/3$ and $c/4$, we get $\xi \approx 3\times 10^2$,  and $\xi \approx7\times 10^2$, respectively. If $W_{\rm p}$ is the energy density of the excited plasma waves, then the average spectral energy density will be
\begin{equation}
W_k=\frac{c^3}{\omega^3_{\rm L}}\frac{W_{\rm p}}{\xi}.
\label{sp-dens}
\end{equation}
Hereinafter in the text, we will use this formula for the spectral energy density of plasma waves in the analysis of the transport equations for the brightness temperatures of radiation.

In the inhomogeneous plasmasphere of an exoplanet, the Langmuir frequency $\omega_{\rm L}(l)$ depends on the coordinates, so the plasma wave at the frequency $\omega_{\rm p}$ can be amplified only on a limited spatial scale $L_{\rm p}$,
\begin{equation}
L_{\rm p}=3L_{\rm n}\frac{v_{\rm T}^2}{\omega_{\rm L}^2}\left(k_{\max}^2-k_{\min}^2\right),
\label{L_p}
\end{equation}
where $L_{\rm n}=\left|n(dn/dl)^{-1}\right|$  is the characteristic scale of plasma inhomogeneity in the  generation region of the plasma waves. Under the assumption $k_{\min}^2<<k_{\max}^2\approx \omega_{\rm L}^2/9v_{\rm T}^2$   the size of the amplification region is
$L_{\rm p}\approx (1/3) L_{\rm n}$ (here, $k_{\max}$  approximately corresponds to the minimum phase velocity of plasma waves
$v_{\rm ph} \approx 3v_{\rm T}$,  at which Landau damping provided by the particles of the main plasma is switched on).

\begin{figure}
	\includegraphics[width=\columnwidth]{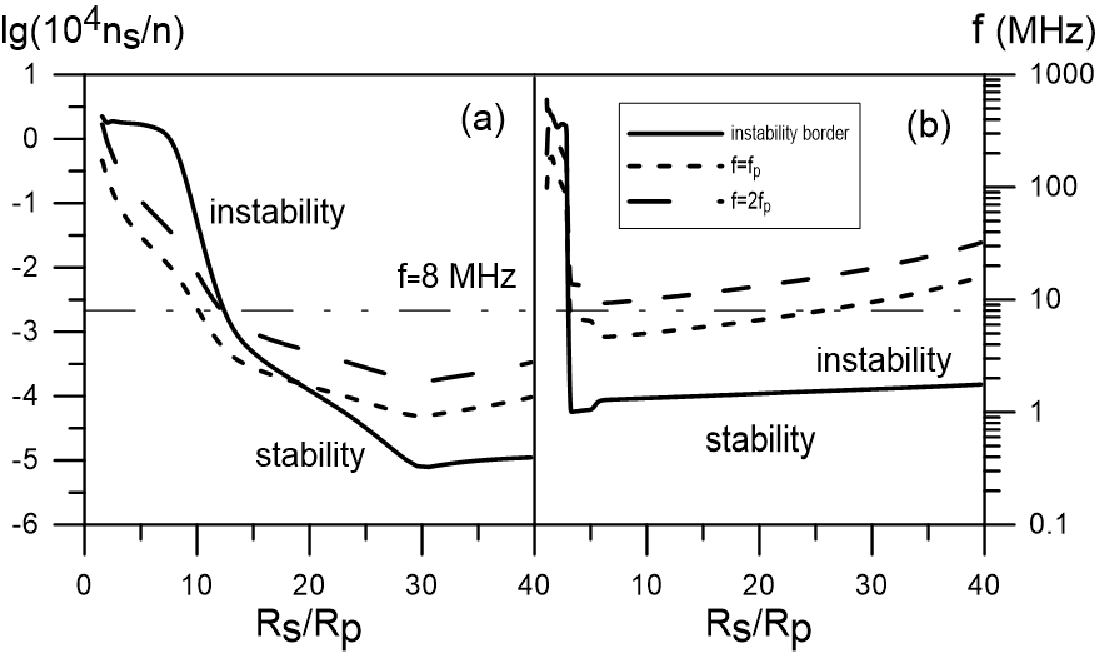}
    \caption{Changes with height of threshold value of energetic electron density, above which the  instability growth rate exceeds the effective frequency of electron-ion collisions (solid line) and of possible frequency of the generated radio emission (dashed lines) at a moderate stellar wind intensity
     (a) and at an intense wind (b).}
    \label{fig2}
\end{figure}

Note also that the necessary condition for the generation of Langmuir waves in the exoplanet's plasmasphere is the excess of the instability grow rate over the effective frequency of electron-ion collisions, $\gamma > \nu_{\rm ei}\approx 50n/T^{3/2}$. This condition depends on the distributions of the plasma density and temperature in the planet's plasmasphere, which, in turn, can change with changes in the activity of the parent star (intensity of the stellar wind flux). Fig.~\ref{fig2} shows
changes with height of threshold value of energetic electron density, above which the  instability growth rate exceeds the effective frequency of electron-ion collisions    and of possible frequency of the generated radio emission  at different stellar wind intensity.

Under normal conditions, registration of radio emission by radio astronomical equipment located on the Earth, is possible at frequencies above the plasma frequency in the ionospheric maximum, $f_{\rm cut} \gtrsim 8$~MHz. This condition limits the position of the plasma source, from which radio emission at the plasma frequency can be detected, to the area close to the surface of the planet, $R \lesssim 10R_{\rm p}$ with moderate wind (Fig.~\ref{fig2}a) and to the areas $R\lesssim3R_{\rm p}$  and $R\gtrsim 25R_{\rm p}$ with intense wind (Fig.~\ref{fig2}b). The situation radically changes when radio emission is generated at the doubled plasma frequency at a high intensity of the stellar wind (Fig. 2b). In this case, the generation of radiation at frequencies above the frequency $f_{\rm cut}$  is possible in the entire plasmasphere.

\section{Conversion of plasma waves into electromagnetic waves in the plasmasphere of HD 189733b}

Scattering of plasma waves with frequency $\omega_{\rm p}$ and wave vector $\vec{k}$ on ions of background plasma with conversion into electromagnetic waves (Rayleigh scattering) leads to the radio emission at a frequency close to the Langmuir one
\begin{equation}
\omega_{\rm t}=\omega_{\rm p}+(\vec{k}_{\rm t}-\vec{k})v_{\rm Ti}\approx \omega_{\rm p}\approx \omega_{\rm L},
\label{sinkhr-1}
\end{equation}
where $\omega_{\rm t}$ and $\vec{k}_{\rm t}$ are the frequency and wave vector of the electromagnetic wave, respectively, $\vec{v}_{\rm Ti}$ is the thermal velocity of equilibrium plasma ions. Nonlinear scattering of plasma waves (Raman scattering) generates radio emission at twice the plasma frequency
\begin{equation}
\omega_{\rm t}(\vec{k}_{\rm t})=\omega^{(1)}_{\rm p}(\vec{k}_1)+\omega^{(2)}_{\rm p}(\vec{k}_2)\approx 2\omega_{\rm p}\approx 2\omega_{\rm L};\ \ \ \vec{k}_1+\vec{k}_2=\vec{k}_{\rm t}.
\label{sinkhr-2}
\end{equation}
The frequency of the electromagnetic wave $\omega_{\rm t}$ is related to the wave vector $\vec{k}_{\rm t}$
\begin{equation}
\omega^2_{\rm t}=\omega^2_{\rm L}+k^2_{\rm t} c^2 ;\ \ \ k_{\rm t}= \frac{\omega_{\rm t}}{c}\sqrt{1-\frac{\omega^2_{\rm L}}{\omega^2_{\rm t}}}.
\label{om-t-disp}
\end{equation}

In this section, we will investigate the contribution of these processes to electromagnetic radiation under the assumption of a sufficiently weak magnetic field in the exoplanet plasmasphere, when the electron gyrofrequency is much lower than the plasma frequency ($\omega_{\rm c} \ll \omega_{\rm L}$).

The transfer equation for the brightness temperature $T_{\rm b}$ of the radio emission is
\begin{equation}
\frac{dT_{\rm b}}{dl}=a-\left(\mu_{\rm N}+\mu_{\rm c}\right)T_{\rm b},
\label{eq_trans}
\end{equation}
where $a$ is the emissivity describing spontaneous emission at the plasma or doubled plasma frequency, $\mu_{\rm N}$ is the emission (absorption) coefficient due to nonlinear processes (induced scattering of plasma waves or the decay of electromagnetic radiation at the second harmonic into two plasma waves), $\mu_{\rm c}$ is the collisional absorption coefficient  (in the radiation generation region $\mu_{\rm c}\equiv\mu_{\rm c}^{\rm in}$ and outside it $\mu_{\rm c}\equiv\mu_{\rm c}^{\rm out}$) given by \citet{Ginzburg(1964)}
\begin{equation}
\mu_{\rm c}=\frac{\omega^2_{\rm L}\nu_{\rm ei}}{\omega^2_{\rm t}v_{\rm g}}.
\label{mu-c}
\end{equation}
In (\ref{mu-c}) $ v_{\rm g}=c\left(1-\omega^2_{\rm L}/\omega^2_{\rm t}\right)^{1/2}$ is the group velocity of electromagnetic waves. The solution to this equation is
\begin{equation}
T_{\rm B}=\frac{a}{\mu_{\rm c}^{\rm in}+\mu_{\rm N}}\left[1-\exp\left(-\tau_{\rm c}^{\rm in}-\tau_{\rm N}\right)\right]\exp\left(-\tau_{\rm c}^{\rm out}\right),
\label{T-B}
\end{equation}
where $\tau_{\rm c}=\int \mu_{\rm c}dl$  determines the value of collisional absorption along the propagation path of radio emission in the generation region 
$\tau_{\rm c}\equiv\tau_{\rm c}^{\rm in}$ and outside it $\tau_{\rm c}\equiv\tau_{\rm c}^{\rm out}$. The brightness temperature is related to the radio emission flux $F$ at a distance $R_{\rm so}$ from the source by the relation \citep{Zheleznyakov(1996)}
\begin{equation}
F=\frac{k_{\rm t}^2\kappa_{\rm B} T_{\rm b}}{(2\pi)^2}\frac{S_{\rm s}}{R_{\rm so}^2},
\label{flux}
\end{equation}
where $S_{\rm s}$ is source area on line of sight.

\subsection{Maser effect in Rayleigh scattering}

When the coefficient of induced scattering $\mu_{\rm N}=\mu{\rm sc}$  is negative and sufficiently large in absolute value, $\left| \mu_{\rm sc}\right|>\mu_{\rm c}^{\rm in}$ the sum $\mu_{\rm sc}+\mu_{\rm c}^{\rm in}$ in equation (\ref{eq_trans}) becomes negative. As a result, an exponential growth $(-\tau_{\rm c}^{\rm in}-\tau_{\rm N}>0)$ of the intensity of electromagnetic radiation occurs, i.e. the maser effect is realized. The efficiency of maser amplification of electromagnetic waves depends on the optical thickness of the source
\begin{equation}
\tau=\left|\tau_{\rm sc}+\tau_{\rm c}^{\rm in}\right|=\int_{0}^{L_{\rm p}}\left|\mu_{\rm sc}+\mu_{\rm c}^{\rm in}\right| dl,
\label{tau}
\end{equation}
which at the corresponding energy of plasma waves can reach sufficiently large values and provide significant fluxes of radio emission emerging from the source. In (\ref{tau}) $L_{\rm p}$ is  determined by formula (\ref{L_p}).

In astrophysical plasma, upon scattering by ions (\ref{sinkhr-1}), the width of the spectrum of plasma waves is, as a rule, greater than the width of the kernel of the integral equation describing the induced conversion. In this case, the scattering of plasma waves into electromagnetic waves has a differential character, when in each act of scattering the frequency of the scattered plasma wave changes by the value $\Delta\omega_{\rm p}<kv_{\rm Ti}$. In this approximation and, assuming an isotropic plasma wave spectrum, the coefficients of spontaneous $a$ and induced scattering $\mu_{\rm N}$ in equation (\ref{eq_trans}) are \citep{Tsytovich(1977)}:
\begin{equation}
a_{\rm sc}\approx\frac{\pi}{36}\frac{\omega_{\rm L}^3 W_k}{v_{\rm g}nv_{\rm T}^2k},
\label{a_sc}
\end{equation}
\begin{equation}
\mu_{\rm sc}\approx -\frac{\pi}{108}\frac{m_{\rm e}\omega_{\rm L}^3}{m_{\rm i}v_{\rm g}n \kappa_{\rm B} Tv_{\rm T}^2k}\frac{\partial}{\partial k}\left(kW_k\right),
\label{mu_sc}
\end{equation}
where $m_{\rm i}$ is the ion mass, $W_k$ is the spectral density of plasma waves, which is related to the plasma wave energy density $W_{\rm p}$ by the relation $W_{\rm p}=\int W_k dk$.

Let us pass in equation (\ref{eq_trans}) from integration over the spatial coordinate $l$ along the propagation of radiation to integration over the wave vector of plasma waves from $k_{\min}$ to $k_{\max}$. Taking into account that $ \omega^2_{\rm L}+3k^2 v_{\rm T}^2=\omega_{\rm p}^2\approx \omega_{\rm t}^2=const$, we obtain the following relationship between $dl$ and $dk$
\begin{equation}
dl=6L_{\rm n}\frac{v_{\rm T}^2}{\omega_{\rm p}^2}kdk.
\label{dl}
\end{equation}
As a result, for the optical thickness of the induced conversion of plasma waves into electromagnetic waves, we obtain
\begin{equation}
\tau_{\rm sc}=\int_{0}^{L_{\rm p}}\mu_{\rm sc}dl\approx-\frac{\pi}{18\sqrt{3}}\frac{m_{\rm e}\omega_{\rm L}\langle v_{\rm ph}\rangle}{m_{\rm i}cv_{\rm T}}wL_{\rm n},
\label{tau_sc}
\end{equation}
where $\langle v_{ph}\rangle =\omega_{\rm p}/\langle k\rangle$ is the average phase velocity of plasma waves, $w= W_{\rm p}/n\kappa T$  is
 the ratio of the plasma wave energy density to the thermal plasma energy density.
 The negative value of $\tau_{\rm sc}$ means that the brightness temperature of the electromagnetic radiation can grow exponentially with increasing plasma wave energy density.
 The optical depth of absorption of electromagnetic waves due to Coulomb collisions in the plasma wave generation region is equal to
\begin{equation}
\tau_{\rm c}^{\rm in}=\int_{0}^{L_{\rm p}}\mu_{\rm c}dl\approx \frac{6}{\sqrt{3}}\frac{v_{\rm T} \nu_{\rm ei}}{c\langle v_{\rm ph}\rangle}L_{\rm n}.
\label{tau_c-in}.
\end{equation}

As noted above, the maser amplification of electromagnetic radiation  occurs when the induced conversion of plasma waves into electromagnetic waves in the source  exceeds the collisional absorption of electromagnetic waves on the plasma wave amplification scale $L_{\rm p}$. This imposes a limitation on the energy density of plasma waves, which must exceed a certain threshold. From the requirement $\left|\tau_sc \right|>\tau_c^{\rm in}$ we obtain
\begin{equation}
w>\frac{108}{\pi}\frac{m_{\rm i}v_{\rm T}^2 \nu_{\rm ei}}{m_{\rm e}\langle v_{\rm ph}^2\rangle \omega_{\rm L}}.
\label{w}
\end{equation}
\begin{figure}
	\includegraphics[width=\columnwidth]{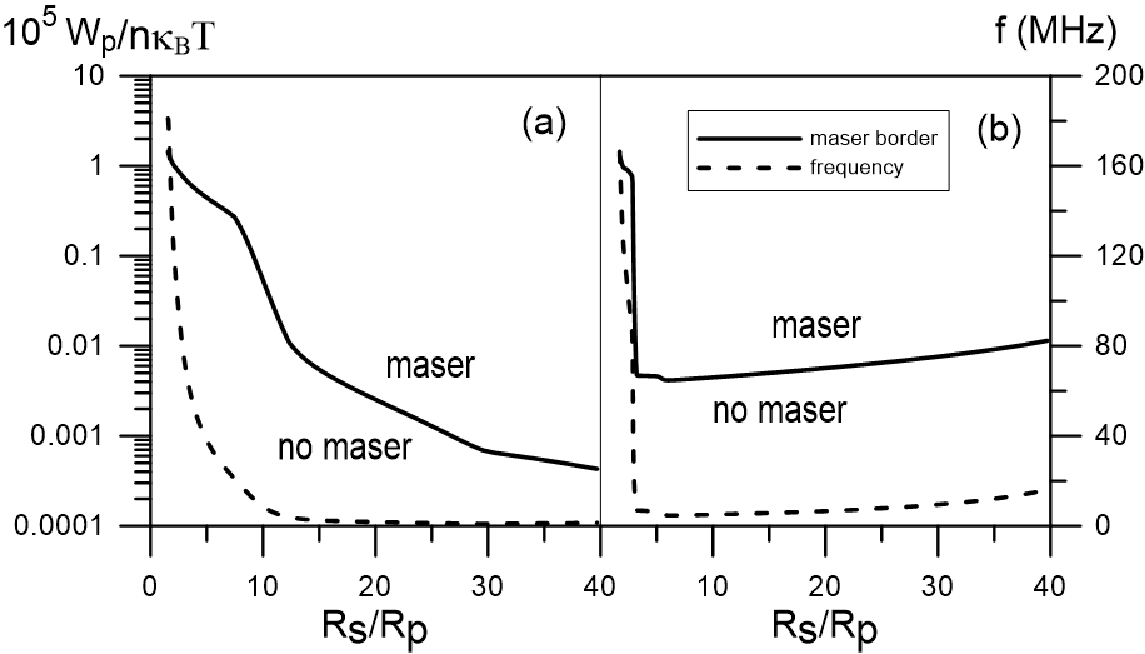}
    \caption{The minimum value of the relative energy density of plasma waves which is necessary to compensate for collisional absorption (solid line) and the frequency of generated radio emission (dashed line) depending on the height of the source in the plasmasphere at moderate wind intensity (a) and at intense wind (b).}
    \label{fig3}
\end{figure}

It follows from Fig.~\ref{fig3} that condition (\ref{w}) is satisfied already at relatively low energy densities of plasma waves, which are significantly less than the values required to generate a radio emission flux of 1~Jy at the Earth (see below).

In the case of the maser effect, $-\tau_{\rm c}^{\rm in}-\tau_{\rm sc}>0$ and $\tau_{\rm c}^{\rm in}\ll |\tau_{\rm sc}|$, the brightness temperature of the radiation emerging from the planet's atmosphere is determined by the formula
\begin{equation}
T_{\rm B}=3\frac{m_{\rm i}}{m_{\rm e}}T\left[\exp (-\tau_{\rm sc})-1\right]\exp (-\tau_{\rm c}^{\rm out}),
\label{T-B-maser}
\end{equation}
where $\tau_{\rm c}^{\rm out}$ is the optical depth of collisional absorption on the way from the source to the observer, to which the main contribution comes from absorption in the explanet's ionosphere outside the region of plasma wave generation and absorption in the stellar wind plasma. From (\ref{flux}), (\ref{tau_sc}) and (\ref{T-B-maser}) we obtain the following expression for the flux of electromagnetic radiation observed on Earth                                                                                                    \begin{multline}
F=3\frac{\kappa_{\rm B} T m_{\rm i}}{c^2 m_{\rm e}}f_{\rm t}^2\frac{S_{\rm s}}{R_{\rm so}^2}\{\exp \left[\frac{\pi}{18\sqrt{3}}\frac{m_{\rm e}\omega_{\rm L}\langle v_{\rm ph}\rangle}{m_{\rm i}cv_{\rm T}}L_{\rm n}w\right]-1\} \\ \exp \left(-\tau_{\rm c}^{\rm out}\right),
\label{F_out}
\end{multline}
where $f_{\rm t}=\omega_{\rm t}/2\pi \approx \omega_{\rm L}/2\pi$ is the electromagnetic wave frequency.

\begin{figure}
	\includegraphics[width=\columnwidth]{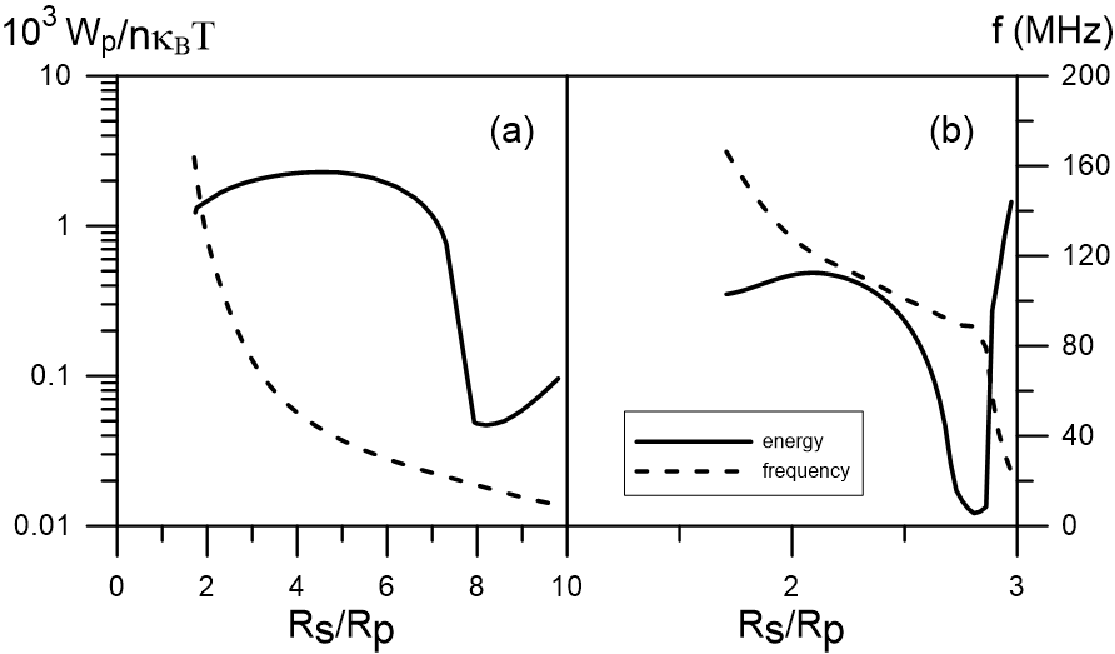}
    \caption{Relative (with respect to the equilibrium plasma energy density) plasma wave energy density required to generate a radio emission flux of 1~Jy on the Earth and its frequency depending on the height of the source in the plasmasphere at moderate wind intensity (a) and at intense wind (b). The upper solid line corresponds to plasma waves with phase velocities equal to $\langle v_{\rm ph}\approx c/3 \rangle$ , the lower - line $\langle v_{\rm ph}\rangle\approx c/4$}
    \label{fig4}
\end{figure}

It follows from Fig.~\ref{fig4} that, depending on the intensity of the stellar wind, the requirements for the energy density of plasma waves change significantly.  In strong winds, the required density is about an order of magnitude lower than in moderate winds, which is associated with a decrease in the extent of the dense part of the exoplanet's plasmasphere.  Fig.~\ref{fig4}b shows the required energy density of plasma waves only in the generation region closest to the exoplanet ($R\lesssim 3R_{\rm p}$). Estimates show that in the more distant ($R\gtrsim 25R_{\rm p}$) region where  generation of radio emission with a frequency above the cutoff frequency $f_{\rm cut}$ is possible, the required relative energy density is significantly higher, $W_{\rm p}/nk_{\rm B} T\gtrsim 5\times 10^{-2}$. The latter is primarily due to the low density of the equilibrium plasma at the periphery of the exoplanet's plasmasphere.

\subsection{Radio emission at double plasma frequency}

As noted above, due to Raman scattering of plasma waves, electromagnetic radiation is generated at the doubled plasma frequency $\omega_{\rm t} (\vec{k}_{\rm t}) \approx 2\omega_{\rm p}$, with the wave vector $\vec{k}_{\rm t}=\vec{k}^{(1)}+\vec{k}^{(2)}$ (\ref{sinkhr-2}). If plasma waves are excited by energetic electrons with velocities $v_{\rm Ts}\ll c$, then $k_{\rm t}\ll k$ and Raman scattering occurs on counterpropagating waves ($\vec{k}^{(1)}\approx -\vec{k}^{(2)}$). In the case of an isotropic spectrum of plasma waves, the coefficients in the transport equation (\ref{eq_trans}) have the form (\cite{Zheleznyakov(1996)})
\begin{equation}
a=a_{\rm cp}\approx\frac{(2\pi)^5}{15\sqrt{3}}\frac{c^3}{\omega_{\rm L}^2\langle
v_{\rm ph}\rangle}\frac{w^2}{\xi^2}nT; \ \ \ \mu_{\rm N}=
\mu_{\rm dc}\approx\frac{(2\pi)^2}{5\sqrt{3}}\frac{\omega_{\rm L}}{\langle v_{\rm ph}\rangle}\frac{w}{\xi},
\label{a-cp}
\end{equation}
where $a_{\rm cp}$ is the coefficient of spontaneous scattering into the second harmonic, the coefficient $\mu_{\rm dc}$ characterizes the absorption of radiation at the second harmonic due to the decay of the electromagnetic wave of the second harmonic into two plasma waves.

The solution of the transport equation (\ref{eq_trans}) has the form (\ref{T-B-maser}), where the coefficients are $a=a_{\rm cp}$, $\tau_{\rm N}=\tau_{\rm dc}=\int_0^{L_{\rm p}}\mu_{\rm dc}dl$. In the case of an optically thick source, $\tau_{\rm dc}+\tau_{\rm c}^{\rm in}\gg 1$, solution (\ref{T-B-maser}) takes the form
\begin{equation}
T_{\rm B}(2\omega)\approx\left(\frac{a_{\rm dc}}{\mu_{\rm dc}+\mu_{\rm c}^{\rm in}}\right)\exp (-\tau_{\rm c}^{\rm out}).
\label{T-B-coupling}
\end{equation}

Further, to estimate the radio emission flux to the second harmonic of the plasma frequency from the exoplanet, we will neglect the collisional absorption $\mu_{\rm c}^{\rm in}$  in the generation region. Estimates show that at the energy density of plasma waves,  $W_{\rm p} \lesssim 10^{-3} n\kappa_{\rm B} T$ (implementation of larger values is hardly possible) this is justified at the distance $R \gtrsim 10R_{\rm p}$ in case of moderate stellar wind intensity and at the distance $R\gtrsim 5R_{\rm p}$ in case of strong wind.

\begin{figure}
	\includegraphics[width=\columnwidth]{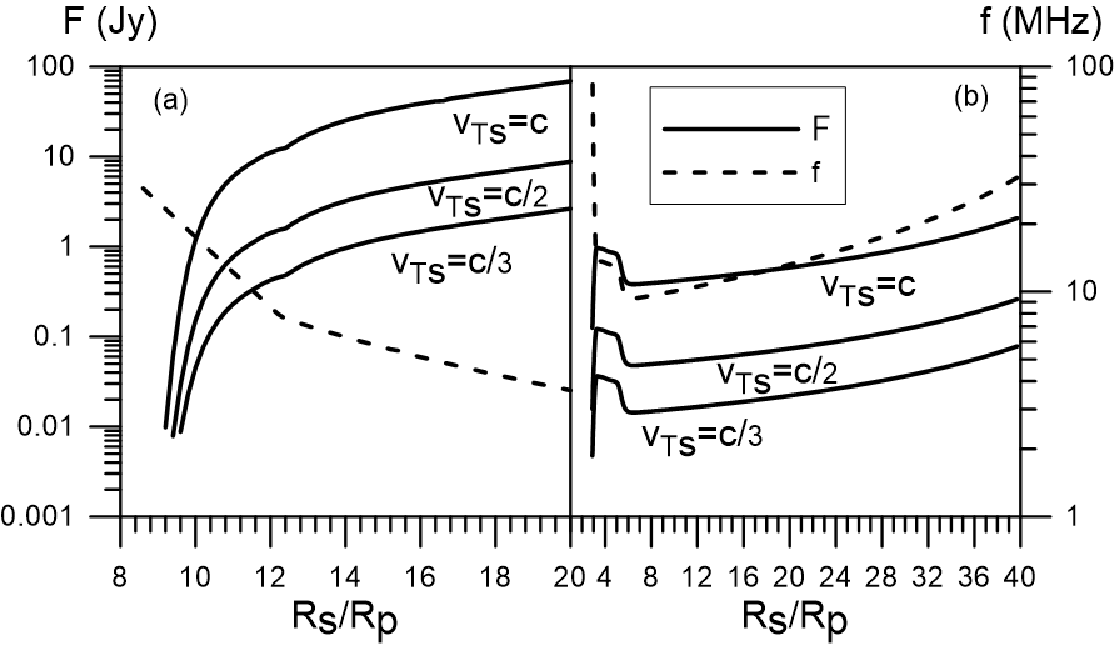}
    \caption{Change in the radio emission flux at the Earth (solid line) and its frequency (dashed line) depending on the height of the source in the case of generation at twice the plasma frequency at moderate wind intensity (a) and at intense wind (b).}
    \label{fig5}
\end{figure}

Fig.~\ref{fig5} shows the radio emission fluxes at the Earth at the doubled plasma frequency and the energy density of plasma waves $W_{\rm p}=10^{-3} n\kappa_{\rm B} T$  for different velocities of energetic electrons in the case of the plasmasphere interacting with the stellar wind of moderate (Fig.~\ref{fig5}a) and high intensity (Fig.!\ref{fig5}b). It was assumed that the area of the radio emission source, located at a distance $R$ from the planet, is equal to the maximum $S=2\pi R^2$.

It can be seen from Fig.~\ref{fig5}a that  the radio emission flux available for observation on the Earth increases sharply, from several milliJansky at a frequency of $30$~MHz to several Jansky at a frequency of $\approx 8$~MHz in the case of a moderate stellar wind. A similar increase in flow is observed in strong winds (Fig.~\ref{fig5}b). However, the flux at the same energy density of the plasma waves is essentially lower. The latter is caused by changes in the density distribution in the plasmasphere.
 In addition, Fig.~\ref{fig5} also illustrates a noticeable increase in the radio emission flux with an increase in the speed of energetic electrons.

\begin{figure}
	\includegraphics[width=\columnwidth]{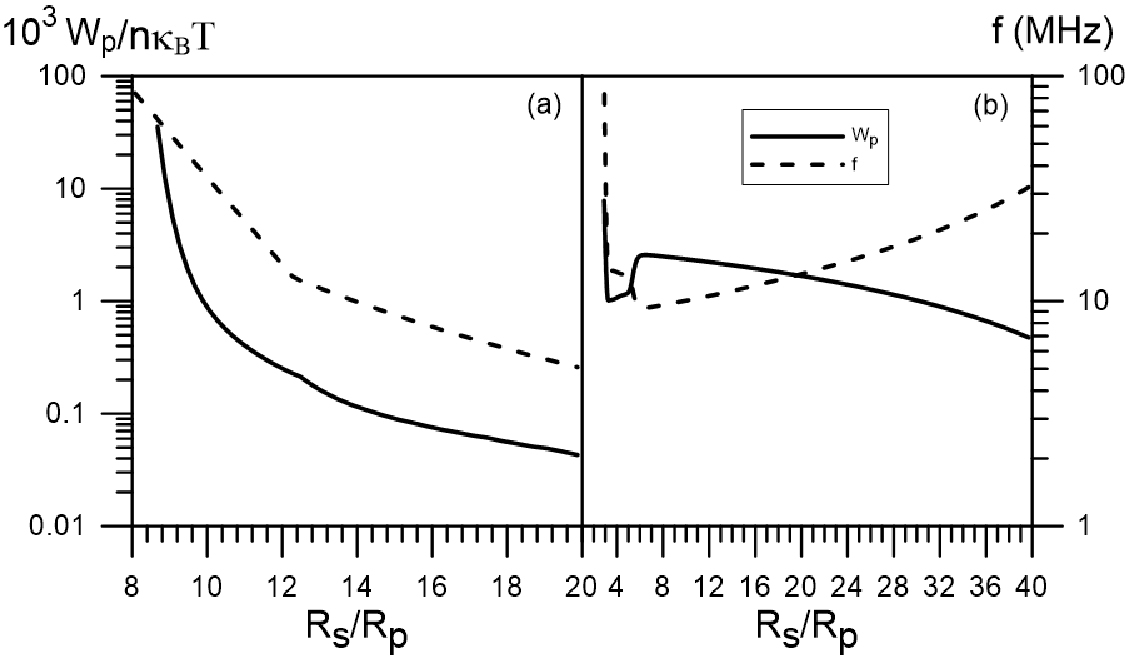}
    \caption{Change in the energy density of plasma waves required to obtain a radio emission flux of 1~Jy at the Earth (solid line) and the frequency of the received radio emission (dashed line) depending on the distance of the source from the surface of the exoplanet in the case of moderate (a) and intense (b) stellar wind.}
    \label{fig6}
\end{figure}

The  energy of plasma waves required to obtain a radio emission flux of 1~Jy at the Earth and the frequency of the received radio emission depending on the distance of the source from the surface of the exoplanet in  a moderate stellar wind, are shown in Fig.~\ref{fig6}a and in a strong wind - in Fig.~\ref{fig6}b.

In the case of a moderate wind, low enough plasma wave energy, $W_{\rm p}\ll 10^{-3}n\kappa_{\rm B} T$ , is necessary to generate a flux of radio emission available for registration on Earth by modern radio telescopes. The frequency of this radio emission is in the range $f_{\rm t}\lesssim 20$~MHz (Fig.~\ref{fig6}a).

It follows from Fig.~\ref{fig6}a that sufficiently low, $W_{\rm p}\ll 10^{-3}, n\kappa_{\rm B} T$ plasma wave energy densities, necessary for the formation of radio emission fluxes accessible for observation by modern radio astronomical means, correspond to the frequency range $f_{\rm t}\lesssim 20$~MHz. An increase in the intensity of the stellar wind (Fig.~\ref{fig6}b) leads to a shift in the frequency range to higher frequencies of $10-30$~MHz, and to increase by about an order of magnitude of the required energy of plasma waves.

\section{Conclusions}

As the study, carried out in this paper, showed, the plasma mechanism for generating radio emission from the exoplanet HD~189733b can be implemented in the planet's plasmasphere at any intensity of the stellar wind from the parent star. However, the efficiency of the plasma mechanism, the conditions for generating radiation, and the possible frequency range of radio emission available for registration on Earth vary depending on the wind intensity. This is due to significant changes in the structure of the plasmasphere and the dependence of the plasma mechanism on the plasma parameters. In particular, when the stellar mass carried away by the wind increases from $10^{11} \mbox{ g s$^{-1}$}$ (moderate stellar wind) to $2 \times 10^{13} \mbox{ g s$^{-1}$}$ (intense stellar wind), a region with dense plasma ($\gtrsim 10^8  \mbox{ cm$^{-3}$}$)  appears concentrated near the exoplanet at a distance of $\lesssim 2.5 R_{\rm p}$. Beyond this region there is a sharp decrease in the plasma concentration by approximately two orders of magnitude with the formation of a ''plateau''. These changes, in turn, affect the efficiency of the plasma mechanism. Thus,  in an intense stellar wind, the energy of plasma waves required to generate the same radio emission flux at the first plasma harmonic is approximately an order of magnitude lower than in a moderate stellar wind. In contrast, when generating at twice the plasma frequency, the required plasma wave energy density is reduced in moderate wind. The latter is due to a noticeable decrease in the frequency of Coulomb collisions outside the generation region and a decrease in the absorption of radio emission in the exoplanet's plasmasphere.
The frequency range of radio emission available for registration on Earth by modern radio astronomical means shifts to higher frequencies with increasing intensity of the stellar wind.  For the stellar wind intensity values $10^{11} \mbox{ g s$^{-1}$}$  and $2 \times 10^{13} \mbox{ g s$^{-1}$}$ considered in the paper, the frequency interval of the radio emission shifts from about $8-20$~MHz in the case of a moderate wind to $10- 30$~MHz in case of intense stellar wind.

In general, our analysis shows that the plasma generation mechanism is capable of generating the radio emission of an exoplanet accessible for registration on Earth under rather mild restrictions on the parameters of the plasmasphere at any intensity of the stellar wind. A change in the intensity of the stellar wind can manifest itself in a shift in the frequency spectrum of the detected radio emission, which will make it possible to use the radio emission characteristics to analyze the activity of the parent star.

\section*{Acknowledgements}

The work was supported by  the Russian Science Foundation under grant No 23-22-00014.

\section*{Data availability}

This study is theoretical and does not use any specific observation data. All data used in the study were derived from published articles. Links
to relevant articles are given in the list of references.

\bibliographystyle{mnras}
\bibliography{Exoplanets_libr_2023}

\label{lastpage}
\end{document}